\documentclass[twocolumn, superscriptaddress, floatfix, showpacs, aps, amsmath, amssymb, amsfonts, pra]{revtex4} 

\usepackage{graphicx}
\usepackage{dcolumn}
\usepackage{bm}
\usepackage{amsmath,amssymb,amsfonts}

\begin{document}

\newcommand{\beq}{\begin{equation}}
\newcommand{\eeq}{\end{equation}}
\newcommand{\beqa}{\begin{eqnarray}}
\newcommand{\eeqa}{\end{eqnarray}}
\newcommand{\br}{\mathbf{r}}
\newcommand{\bx}{\mathbf{x}}
\newcommand{\bxp}{\mathbf{x'}}
\newcommand{\GHF}{G_{\rm{HF}}}

\newcommand{\psih}{\hat{\psi}}
\newcommand{\psihd}{\hat{\psi}^{\dagger}}
\newcommand{\hpsid}{\hat{\psi}^{\dagger}}
\newcommand{\hpsi}{\hat{\psi}}

\newcommand{\psin}{\Psi_0^N}
\newcommand{\psinmu}{\Psi_i^{N-1}}

\newcommand{\bra}[1]{\langle #1|}l.
\newcommand{\ket}[1]{|#1\rangle}
\newcommand{\braket}[2]{\langle #1|#2\rangle}
 \arraycolsep=1.4pt

\title{Levels of self-consistency in the $GW$ approximation\footnote{See J. Chem. Phys. {\bf 130} for the published version.}}
\author{Adrian Stan}
\affiliation{Department of Physics, Nanoscience Center, FIN 40014, University of Jyv\"askyl\"a,
Jyv\"askyl\"a, Finland}
\affiliation{European Theoretical Spectroscopy Facility (ETSF)}
\affiliation{Rijksuniversiteit Groningen, Zernike Institute for Advanced Materials, Nijenborgh 4, 9747AG Groningen, The Netherlands.}
\author{Nils Erik Dahlen}
\affiliation{Rijksuniversiteit Groningen, Zernike Institute for Advanced Materials, Nijenborgh 4, 9747AG Groningen, The Netherlands.}
\author{Robert van Leeuwen}
\affiliation{Department of Physics, Nanoscience Center, FIN 40014, University of Jyv\"askyl\"a,
Jyv\"askyl\"a, Finland}
\affiliation{European Theoretical Spectroscopy Facility (ETSF)}
\date{\today}

\begin{abstract}
We perform $GW$ calculations on atoms and diatomic molecules
at different levels of self-consistency and investigate the effects of self-consistency on total energies, 
ionization potentials and on particle number conservation. 
We further propose a partially self-consistent $GW$ scheme in which we keep the correlation part of the self-energy fixed
within the self-consistency cycle.
This approximation is compared to the fully self-consistent $GW$ results and to the $G W_0$ and the
$G_0W_0$ approximations. Total energies, ionization potentials and two-electron removal
energies obtained with our partially self-consistent $GW$ approximation are in excellent
agreement with fully self-consistent $GW$ results while requiring only a fraction of the computational effort.
We also find that self-consistent and partially self-consistent schemes 
provide ionization energies of similar quality as the $G_0W_0$ values but yield better total energies and
energy differences. 
\end{abstract}

\pacs{31.15.-p,31.15.xm,71.15.-m}

\keywords{many-body Green's function, GW approximation, self-consistency, } 

\maketitle

\section {Introduction}

Green function methods  \cite{fetterwalecka,rungegrossbook} have been very succesful
in the description of various properties of many-electron systems,
ranging from atoms and molecules to solids \cite{aryasetiawan98, albur}.
Within the Green function approach, these properties are completely determined by
the self-energy operator $\Sigma$, which incorporates all the effects of exchange and correlation in a many-particle system \cite{fetterwalecka}.
One of the most widely used approximations to the self-energy is the $GW$ approximation ($GW$A) \cite{hedin65}. 
In the $GW$A, the 
self-energy operator has the simple form $\Sigma=-GW$, where $G$ is the Green function that describes the propagation of particles and 
holes in the system, and $W$ is the dynamically screened interaction. This quantity
 describes how the bare interaction $v$ between electrons is modified due to the presence 
of the other electrons and appears as a renormalized interaction in terms of Feynman diagrams. 
In extended systems the screened interaction is much weaker than the bare interaction, and therefore it is much more 
natural to expand the self-energy in terms of the screened interaction than in terms of the bare interaction. The lowest order in this expansion~\cite{hedin65} is the $GW$A. \\
Calculations within the $GW$A are usually done in two steps. First, a density functional theory (DFT)~\cite{DreizlerGross} calculation is performed and the DFT orbitals and eigenvalues are used to construct a first guess $G_0$, for the Green function and a first guess $W_0$, 
for the screened interaction. In a second step, the self-energy $\Sigma=-G_0 W_0$ is constructed
and the Dyson equation is solved for the Green function. In principle, this new Green 
function should be used to calculate a new self-energy and this process should be iterated to self-consistency \cite{hedin65}. However, one usually
stops after the first iteration. 
The corresponding approximation for the Green function is known as the
 $G_0 W_0$ approximation and has become one of the most accurate methods for the 
 calculation of spectral properties and band gaps of solids \cite{aryasetiawan98, albur}.
One reason for not going beyond the first iteration of the $G_0W_0$ method is the large
computational cost involved. There are further indications that a full self-consistent
solution would worsen the spectral properties as a consequence of a cancellation
between dressing of Green functions and vertex corrections~\cite{mahan94}. This was 
investigated for the
electron gas~\cite{holm98} and the Hubbard model~\cite{schindlmayr98hubbard}. 
However, this problem has not been investigated in detail for real systems mainly due to the 
computational cost involved.\\
The $G_0W_0$ approximation has, however, two unsatisfactory aspects. The first aspect is related 
to the satisfaction of conservation laws. Baym \cite{baym62} has shown that the self-energy expressions 
that can be obtained as a functional derivative of a functional $\Phi [G]$ of the Green function, 
i.e. $\Sigma=\delta \Phi / \delta G $, have the important property that 
they lead to conserving many-body approximations. These approximations obey basic conservation laws, like the ones for
particle number, momentum, angular momentum and energy. The $GW$A is one of these conserving 
schemes \cite{baym61,kohnmode, psipaper}. 
However, the $\Phi$-derivable approximations are only conserving when the Dyson equation for the Green function is solved 
fully self-consistently. A lack of full self-consistency will generally result in a violation of the conservation laws. 
For this reason the use of conserving approximations, such as $GW$, is crucial in obtaining a  correct description of transport phenomena within a nonequilibrium Green function approach \cite{thygesen08PRB, thygesen07, thygesen08PRL, hybertsen08a, myohanen08}.
Since it is one of our research goals to study quantum transport, it will be necessary to consider the fully self-consistent $GW$ (SC-$GW$) approximation~\cite{holm98, schindlmayr98PRL, ku02, delaney04, holm99, garcia-gonzalez01, barth04}.\\
A second unsatisfactory aspect of nonself-consistent schemes, such as $G_0 W_0$, is that the values of the observables depend on the way they are 
calculated. For instance, the total energy can be calculated in different ways from the Green function 
and the self-energy: using the Galitskii-Migdal formula~\cite{gm58}, a coupling constant integration~\cite{holm99}, a Luttinger-Ward expression ~\cite{dahlen06, xipaper, psipaper, LW2Deg}
or various other expressions. For nonself-consistent calculations all these expressions 
lead to different results and therefore to ambiguity in the value of the energy.
It was, however, demonstrated in the work of Baym \cite{baym62} that self-consistent $\Phi$-derivable 
approximations are not only conserving but also have the property that
all the various ways in which the observables are calculated provide the same result.
This is another motivation for considering fully self-consistent many-body schemes.\\
We can therefore conclude that self-consistency is important to obtain conserving
and unambiguous results. However,
the large computational cost of self-consistent schemes makes them unattractive for the calculation of the properties
of large and extended systems. In order to lower the computational effort 
it is possible to use partial self-consistency which may result in a less severe violation
of conservation laws. One can, for instance, keep the screened interaction fixed during iteration of 
the Dyson equation. This leads to a scheme that can be shown to still conserve the particle number and 
that has been tested on the electron gas \cite{vonBarthHolm, garcia-gonzalez01}. 
Another approach in which the self-consistency is constrained is the so-called quasi-particle
self-consistent $GW$ (QS$GW$) method~\cite{Kotani:SSC02,Faleevetal:PRL04,Schilfgaardeetal:PRL06,Faleevetal:PRB06}. In this approach a frequency independent self-energy
of $GW$-form is constructed and used to solve a quasi-particle equation from which the Green function and
the screened interaction are constructed iteratively. Due to the Hermitian nature of the self-energy 
the method leads to an orthonormal set of quasi-particle states and thereby restricts the form
of the Green function and the screened interaction.  This method has been succesful in improving the
$G_0 W_0$ band gaps and band widths for a large range of solids~\cite{Schilfgaardeetal:PRL06}. One could further 
consider similar other approximations within a quasi-particle framework~\cite{brunevaletal:PRB06}. Such
approximations have been shown to improve the band structure when local density approximation is a poor 
starting point. These methods are, however, not $\Phi$-derivable and are in general not conserving.
Extending methods based on quasi-particle equations 
to the time-dependent case is not as straightforward as for the SC-$GW$, $GW_0$ and $G_0W_0$ methods,
which are instead based on an equation of motion for the Green function. 
For the same reason the computational schemes used in 
this paper (which aims at an extension to the time-dependent case) 
would need to be modified in order to do QS$GW$ calculations. We therefore did not consider
the QS$GW$ method in this work.
However, we propose another partially self-consistent scheme which is computationally cheaper than
the $GW_0$ method.  In this approximation the correlation part of 
the self-energy is fixed during the iteration cycle while only the Hartree and exchange parts are 
updated self-consistently. In this paper we investigate this approximation and other $GW$ schemes at different levels of self-consistency and test them on atoms and diatomic molecules.
We also present in more detail the computational method behind the self-consistent $GW$ calculations 
that we described briefly in an earlier Letter~\cite{astan06}. 
The paper is divided as follows: In Sec.~\ref{theory} we briefly present the 
general formalism and in Sec.~\ref{SGW-LA} we describe in detail the $GW$ approximation at different 
levels of self-consistency.
We then present in Sec.~\ref{computational} the details of our computational 
procedure. Finally, in Sec.~\ref{results}, we will discuss the results obtained with the $GW$A at different 
levels of self-consistency for atoms and some diatomic molecules. 
These systems are well-suited to test the $GW$ at 
different levels of self-consistency, but we are ultimately interested in applications in quantum transport 
theory for molecules attached to macroscopic leads. In such applications the long range screening 
effects, as incorporated in the $GW$A, are important. The investigations in this paper are a first step in this
direction and aim to get further insight into various aspects of the $GW$A that are relevant in quantum transport
theory.

\section{General formalism} 
\label{theory}

We study finite many-particle systems using the Matsubara formalism~\cite{fetterwalecka, matsubara55}
which can easily be extended to a nonequilibrium version of the theory~\cite{keldysh65, danielewicz, wagner91}.
We consider a many-body system in thermal equilibrium 
at a temperature $T$ and chemical potential $\mu$, and with the
Hamiltonian (in second quantization~\cite{fetterwalecka})
\begin{eqnarray}
&&\hat{H}=\int\; d\mathbf{x}\;\hat{\psi}^\dagger(\mathbf{x}) h(\mathbf{r}) \hat{\psi}(\mathbf{x}) +  \nonumber \\
&&+\frac{1}{2}\int\int\; d\mathbf{x_1}  d\mathbf{x_2} \hat{\psi}^\dagger(\mathbf{x_1})\hat{\psi}^\dagger(\mathbf{x_2})v(\br_1,\br_2)\hat{\psi}(\mathbf{x_2})\hat{\psi}(\mathbf{x_1}). 
\end{eqnarray} 
Here $\bx=(\br,\sigma)$ denotes the space- and spin coordinates. The two-body 
interaction $v$ is taken to be of Coulombic  form $v(\br_1,\br_2)=1/|\mathbf{r_1}-\mathbf{r_2}|$. We use atomic units 
$\hbar=m=e=1$ throughout this paper. The single particle part of the Hamiltonian $h(\mathbf{r})$ has the explicit form
\begin{equation}
h(\mathbf{r})=-\frac{1}{2}\nabla^2+w(\mathbf{r}) - \mu,
\label{eq:h0}
\end{equation}
where $w(\mathbf{r})$ is the external potential and where we absorbed the chemical potential $\mu$ into $h$.
The equilibrium expectation value of an operator $\hat{O}$ in the grand canonical ensemble is then given by
\beq
\langle \hat{O} \rangle=\text{Tr}\,\{{\hat{\rho}\hat{O}}\},
\label{eq:expval}
\eeq
where 
$\hat{\rho}=e^{-\beta\hat{H}} / \text{Tr}\,e^{-\beta\hat{H}}$
is the statistical operator, $\beta=1/k_BT$ the inverse temperature and $k_B$ is the Boltzmann constant. 
The trace is taken over all states in Fock space~\cite{fetterwalecka}.
The Green function is then defined as
\beqa
 G(\bx \tau_1,\bx' \tau_2)
 = &-& \theta(\tau_1-\tau_2) \langle \hat{\psi}_H(\bx \tau_1)\hat{\psi}_H^\dagger(\bx' \tau_2) \rangle  \nonumber \\
 &+& \theta(\tau_2-\tau_1) \langle \hat{\psi}_H^\dagger(\bx'\tau_2)\hat{\psi}_H(\bx \tau_1) \rangle,
\eeqa
where we define the Heisenberg form of the operators in this equation to be 
$\hat{O}_H=e^{\tau \hat{H}} \hat{O} e^{-\tau \hat{H}}$.
Since the Hamiltonian is time-translation invariant, the equilibrium Green function 
  only depends on the difference between the time coordinates: $G(\bx\tau_1,\bx'\tau_2)=G(\bx,\bx';\tau_1-\tau_2)$. 
The Green function satisfies the equation of motion
\begin{eqnarray}
&&\Big[ - \partial_{\tau} - h(\mathbf{r}) \Big] G({\bf x}, {\bf x'}; \tau)= \nonumber \\
&&= \delta ( \tau ) \delta ({\bf x} - {\bf x'}) + \nonumber \\ &&\int_{0}^{\beta} d \tau_1 \int d {\bf x}_1 \Sigma[G] ({\bf x}, {\bf x}_1; \tau-\tau_1) G({\bf x}_1, {\bf x'}; \tau_1),
 \label{eq:motion} 
\end{eqnarray}
where the self-energy $\Sigma[G]({\bf x}, {\bf x'}; \tau)$ 
incorporates the many-body interactions of the system.
The self-energy can be approximated
with the usual diagrammatic methods~\cite{fetterwalecka,rungegrossbook}.
Since $\Sigma [G]$ is a functional of the Green function Eq.(\ref{eq:motion}) must be solved 
self-consistently. The self-energy is usually split into a Hartree part and an exchange-correlation 
part, according to
\beq
\Sigma [G](\bx_1,\bx_2;\tau) = \delta (\tau) \delta(\bx_1-\bx_2) v_H (\br_1)
+ \Sigma_{\textrm{xc}} [G](\bx_1,\bx_2;\tau),
\label{eq:self}
\eeq
where the Hartree potential is defined as the potential due to the electron charge by
\beq
v_H (\br) = \int d\bx' n(\bx')v(\br,\br'),
\eeq
where we introduced the electron density
\beq
n(\bx) = \lim_{\eta \rightarrow 0} G(\bx,\bx;-\eta).
\eeq
The main task is now to find an approximation for this exchange-correlation part $\Sigma_{\textrm{xc}}$ of the self-energy and to solve Eq.(\ref{eq:motion}).
We convert Eq.(\ref{eq:motion}) to integral form~\cite{dahlen05b}
\beqa
\lefteqn{ G(\bx_1,\bx_2; \tau) = G_0 (\bx_1,\bx_2; \tau) } 
\nonumber \\
&& + 
\int_0^\beta d\tau_1  d\tau_2 \int d\bx_3 d\bx_4 G_0 (\bx_1,\bx_3;\tau-\tau_1) \nonumber \\
&\times& (\Sigma [G](\bx_3,\bx_4;\tau_1-\tau_2) -\delta (\tau_1-\tau_2) \Sigma_0 (\bx_3,\bx_4)) \nonumber \\
&\times& G(\bx_4,\bx_2;\tau_2).
\label{eq:Dyson}
\eeqa
Here we introduced a static reference self-energy $\Sigma_0$
and a reference Green function $G_0$ which is defined by the equation
\beqa
\lefteqn{ \Big[ - \partial_{\tau} - h(\mathbf{r}) \Big] G_0({\bf x}, {\bf x'}; \tau)=} \nonumber \\
&=& \delta ( \tau ) \delta ({\bf x} - {\bf x'}) +  
 \int d {\bf x}_1 \Sigma_0 ({\bf x}, {\bf x}_1 ) G_0({\bf x}_1, {\bf x'}; \tau).
 \label{eq:motion0} 
\eeqa
In practice we solve first Eq.(\ref{eq:motion0}) for $G_0$ and then we solve Eq.(\ref{eq:Dyson}) for $G$. 
It is clear from Eq.(\ref{eq:motion}) that a fully self-consistent solution of Eq.(\ref{eq:Dyson}) does not
depend on the reference Green function $G_0$.
In this work we choose for $\Sigma_0$ a Hartree-Fock (HF) or a density functional self-energy.
In the first case 
$\Sigma_0=v_H [G_0] + \Sigma_x [G_0]$, consisting of Hartree and exchange parts, whereas in the second case
 $\Sigma_0=\delta (\bx-\bx')v_{\textrm{Hxc}}[G_0](\bx)$, where $v_{\textrm{Hxc}}(\bx)$ is the sum of the Hartree and the exchange-correlation potential~\cite{DreizlerGross}.\\
\begin{figure}
\includegraphics[width=8.6cm]{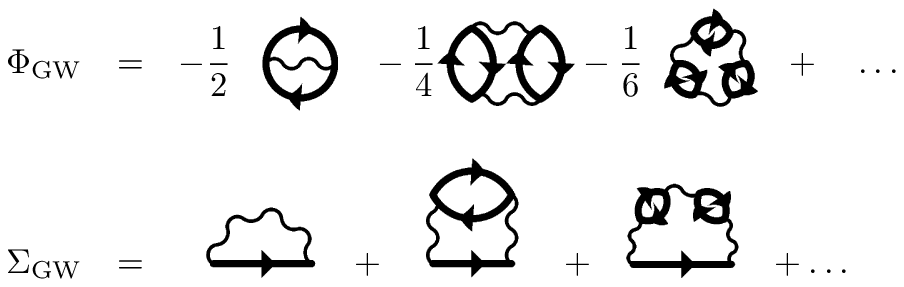}
\caption{ The $GW$ self-energy $\Sigma$ is the functional 
derivative of a functional $\Phi[G]$.}
\label{fig1:gwphi} 
\end{figure}
From the Green function several observables can be calculated.
To calculate the total energy $E=T+V_{ne}+U_0+U_{\rm{xc}}$ we use the fact that the exchange-correlation part
$U_{\rm{xc}}$ of
the interaction energy is given by~\cite{fetterwalecka,rungegrossbook}
\beq
U_{\rm{xc}}=\frac{1}{2} \int_0^\beta d \tau  \int d\bx_1 \int d\bx_2  \Sigma_{\textrm{xc}}(\bx_1,\bx_2;-\tau) 
G(\bx_2,\bx_1;\tau) \label{eq:uxc}.
\eeq
The kinetic energy $T$, the nuclear-electron attraction energy $V_{ne}$, and the Hartree energy $U_0=1/2\int d\br d\br' n(\br)v(\br,\br')n(\br')$ can all be calculated directly from the Green function.
To calculate the ionization potentials from the Green function
we used the extended Koopmans theorem~\cite{katrieldavidson, dayEKT75, smithEKT75, sundholmEKT93, morrison_ayersEKTBe95}, a short derivation of which is given in Appendix~\ref{A2}.

\section{The GW approximation at different levels of self-consistency}
\label{SGW-LA}
\subsection{Fully self-consistent GW}

Within the $GW$A the exchange-correlation part of the self-energy has the explicit form \cite{hedin65, hedin68ei, rexPRL95} 
\beq
\Sigma_{\textrm{xc}}(\bx_1, \bx_2 ; \tau)= - G (\bx_1, \bx_2 ;\tau ) W(\bx_1, \bx_2; \tau),
\label{eq:gw}
\eeq
in which $W$ is a dynamically screened interaction corresponding to an infinite summation of bubble 
diagrams (see Fig.~\ref{fig1:gwphi}). From this figure we see that this self-energy
is given as a functional derivative of a 
functional $\Phi [G]$ with respect to $G$ and hence represents a conserving approximation~\cite{baym62}.
From the diagrammatic structure we see that the screened potential $W$
satisfies the equation 
\begin{eqnarray}
&&W(\bx_1, \bx_2 ;\tau )= v(\br_1,\br_2)\delta(\tau)+ \nonumber \\  
&&+\int d\bx_3 d\bx_4 \int_0^{\beta} d\tau' v(\br_1,\br_3)P(\bx_3, \bx_4; \tau-\tau')W(\bx_4, \bx_2 ; \tau'),
\label{eq:ww}
\end{eqnarray}
where $v$ is the bare Coulomb interaction and $P$ is the irreducible polarization
\beq
P(\bx_1, \bx_2 ; \tau)=  G(\bx_1, \bx_2 ;\tau)G(\bx_2, \bx_1;-\tau).
\label{eq:pp}
\eeq
The problem is now completely defined. Equations (\ref{eq:ww}) and Eq.(\ref{eq:pp}) need to be
solved self-consistently together with Eqs.(\ref{eq:gw}), (\ref{eq:self}) and (\ref{eq:Dyson}). 

\subsection{The $G_0 W_0$ and $GW_0$ approximations}
\label{ssec:gw0}

The $G_0 W_0$ approximation, as mentioned before, is obtained from a single iteration of the Dyson equation Eq.(\ref{eq:Dyson}),
starting from a refence Green function $G_0$. 
For this approximation the 
self-energy is given as $\Sigma_{\textrm{xc}} [G_0] = -G_0 W_0$ where $W_0$ is calculated by
inserting $G_0$ into Eq.(\ref{eq:pp}) and solving Eq.(\ref{eq:ww}) with this irreducible polarization.
The Dyson equation (\ref{eq:Dyson}) is then solved with this self-energy to obtain an improved Green function $G$ 
from which spectral properties are calculated. In principle one should insert this Green function into
the self-energy and solve the Dyson equation again for a new Green function. This procedure should be continued
until self-consistency is achieved, but this is rarely done in practice for the reasons mentioned in
the introduction.\\ 
We further consider a partially self-consistent scheme in which we write the self-energy as 
$\Sigma_{\textrm{xc}}[G, G_0]=- GW_0$, where the Green function $G$ is determined
fully self-consistently by repeated solution of the Dyson equation and where $W_0$ is
calculated from $G_0$ in the same way as for the $G_0W_0$ approximation.
This reduces the computational cost considerably as it avoids the self-consistent calculation of
the screened interaction $W$. 
The corresponding approximation is known as the $G W_0$ approximation~\cite{vonBarthHolm, fortmann08}.  This approximation
was shown to be number conserving by Holm and von Barth~\cite{holmthesis} for the case of homogeneous
systems. More precisely they derived that the $GW_0$ approximation satisfies the Hugenholtz-van Hove 
theorem~\cite{HugenholtzvanHove} for the homogeneous electron gas. However, one can readily derive the number 
conserving property for the inhomogeneous and time-dependent case. This requires nonequilibrium 
Green functions in the proof, but this extension is straightforward~\cite{dahlenKiel324}. If we regard $W_0$ as a given potential
(albeit nonlocal in space and time), it is clear that 
$\Sigma=\delta \Phi/\delta G$ for $\Phi [G,W_0]=-1/2 \mbox{tr} GGW_0$,
where the trace denotes integration over space-time variables.
Since this $\Phi$ is invariant under gauge transformations (the phases cancel
at each vertex of $\Phi$), we can follow the proof of Baym~\cite{baym62} and derive that $GW_0$ is 
particle conserving. However, for time-dependent and inhomogeneous systems $W_0$ is not invariant under 
spatial and time-translations, unlike the bare interaction $v$ that usually appears in the functional 
$\Phi [G]$. Therefore the $GW_0$ approximation will not be momentum or energy conserving.

\subsection{The $GW_{\textrm{fc}}$ approximation}
\label{ssec:gwfc}

The most time-consuming part of the $GW_0$ calculation is the evaluation of the 
correlation part of the self-energy which is nonlocal in time. 
We therefore propose another partial self-consistent scheme
in which we only evaluate the time-local Hartree and exchange parts of the self-energy in
a self-consistent manner.
We therefore split the self-energy as follows
\beq
\Sigma [G,G_0]= \Sigma^{HF}[G] + \Sigma_{\textrm{c}} [G_0 ] .
\label{eq:GWfc}
\eeq
The first term in this equation represents
the Hartree-Fock part of the self-energy
\beq
\Sigma^{HF}[G]=v_{\textrm{H}}[G] + \Sigma_{\textrm{x}}[G] ,
\label{eq:sigmaHF}
\eeq
which consists of a Hartree part and an exchange part $\Sigma_{\textrm{x}}[G]=-Gv$.
The last term in Eq.(\ref{eq:GWfc}) represents the correlation part of the self-energy and
has the explict form
\beq
\Sigma_{\textrm{c}}[G_0] =  -G_0(W_0-v),
\eeq
where $W_0$ is calculated from $G_0$ in the same way as for the $G_0W_0$ approximation.
The approximation for the self-energy of Eq.(\ref{eq:GWfc})  will be denoted as the
$GW_{\textrm{fc}}$ approximation (where fc stands for fixed correlation). 
This approximation is not conserving but, as we will see later, nevertheless produces observables in very 
close agreement with those obtained from a fully SC-$GW$ calculation.

\section{Computational method}
\label{computational}

\subsection{Numerical solution of the Dyson equation}
\label{numericaldyson}
In the following, we will describe the computational methods that we employed for calculating the Green function 
and the screened interaction $W$.
We consider the case of spin-unpolarized systems where the Green function has the form
\beq
G(\bx , \bx' ; \tau)=\delta_{\sigma \sigma'} G(\br, \br' ;\tau).
\label{eq:Green}
\eeq
 The calculations are carried out using a set of basis functions such that
the spin-independent part of the Green function is expressed as
\beq
G(\br,\br';\tau)=\sum_{ij} G_{ij}(\tau) \phi_i(\br) \phi_j^*(\br').
\label{eq:Greenbasis}
\eeq 
The basis functions $\phi_i$ 
are represented as linear combinations of Slater functions 
$\psi_i(\br)=r^{n_i-1}e^{-\lambda_i r}Y_{l_i}^{m_i}(\Omega)$ which are centered on the different nuclei
and are characterized by quantum numbers $(n_i,l_i,m_i)$ and an exponent $\lambda_i$. In these expressions 
and $Y_{l_i}^{m_i}(\Omega)$ are the usual spherical harmonics. 
The molecular orbitals $\phi_i$ and eigenvalues $\epsilon_i$ are obtained from a Hartree-Fock or DFT Kohn-Sham
calculation in this basis.
The particle number $N$ is determined by 
the chemical potential.
Since we consider closed shell systems we have $N/2$ doubly occupied HF or Kohn-Sham levels $\epsilon_i$ 
(some of which may be degenerate).
We therefore choose $\mu$ such that
 $e_i=\epsilon_i - \mu < 0$ for $i \leq N/2$ and  $e_i > 0$ for  $i > N/2$.
In the zero-temperature limit (we used $\beta=100$) the observables are insensitive to the value of $\mu$, provided
$\epsilon_{N/2} <\mu < \epsilon_{N/2+1}$. 
The reference Green function $G_0$ corresponding to the Hamiltonian $h_0+ \Sigma_0$
(either HF or DFT) is diagonal in the basis $\{\phi_i\}$ {\it i.e.}  in matrix form we have 
$G_{ij,0}(\tau)= \delta_{ij} G_{i,0} (\tau)$, where
\beq
G_{i,0} (\tau) = \theta (\tau) (n (e_i)-1) e^{-e_i \tau} + \theta (-\tau) n (e_i) e^{-e_i \tau},
\eeq
and $n(e_j)=(e^{\beta e_j}+1)^{-1}$ is the Fermi-Dirac distribution. 
The Dyson equation of Eq.(\ref{eq:Dyson}) in basis representation has the form
\beqa
\lefteqn{ G(\tau) =G_0(\tau) + } \nonumber \\
&& \int_0^\beta d\tau' \int_0^\beta d\tau'' G_0(\tau-\tau')
\Sigma^c[G, G_0](\tau'-\tau'')G (\tau''), 
\label{eq:alltau}
\eeqa
where we denote
\beq
\Sigma^c[G, G_0](\tau)=\Sigma [G](\tau)-\delta(\tau)\Sigma_0 [G_0],
\label{eq:self-energy-c}
\eeq
and where all quantities are matrices.
Since in the limit $\tau \rightarrow 0^-$, $G$ yields the density matrix, it is convenient to solve the 
Dyson equation for negative $\tau$--values. We therefore rewrite Eq.~(\ref{eq:alltau}) as
\beqa
\lefteqn{ G_{ij}(\tau)=\delta_{ij}G_{i,0}(\tau) +} \nonumber \\
&&\sum_k \int_{-\beta}^0 d\tau_1\int_{-\beta}^0 d\tau_2 G_{i,0}(\tau-\tau_1)\Sigma_{ik}^{c}(\tau_1-\tau_2)G_{kj}(\tau_2),
\label{eq:dysonFIN}
\eeqa
with $\tau\in[-\beta,0]$ where we changed variables $\tau_1 = \tau'-\beta$,
$\tau_2 = \tau''-\beta$,
 and used $G_0 (\tau)=-G_0(\tau+\beta)$ with the same relation for $G$~\cite{dahlen05b}. 
We now discretize Eq.~(\ref{eq:dysonFIN}) using a trapezoidal rule on a time grid 
$(\tau^{(0)}=0, \tau^{(1)}\ldots, \tau^{(m)}=-\beta)$.
Since the Green functions behave exponentially near the endpoints of the imaginary time interval $[-\beta,0]$,
we used a uniform power-mesh~\cite{ku02}. We briefly describe this mesh in Appendix~\ref{A2}.
The discretized version of Eq.~(\ref{eq:dysonFIN}) attains the form
\beqa
\lefteqn{\delta_{ij}G_{i,0}(\tau^{(p)}) = } \nonumber \\
&& \sum_{k,q}\left[\delta_{ik}\delta_{pq}-\frac{\Delta \tau^{(q)}}{2}Z_{ik}(\tau^{(p)},\tau^{(q)}) \right]G_{kj}(\tau^{(q)}),
\label{eq:discrt}
\eeqa
where we defined $Z_{ik}$ as
\beq
Z_{ik}(\tau^{(p)},\tau^{(q)})= \int_{-\beta}^0 d\tau G_{i,0}(\tau^{(p)}-\tau)\Sigma_{ik}^{c}(\tau-\tau^{(q)}).
\label{eq:convolution}
\eeq
The time steps are positive, where $\Delta\tau^{(q)}=\tau^{(q-1)}-\tau^{(q+1)}$ except at the
endpoints where
 $\Delta\tau^{(0)}=\tau^{(0)}-\tau^{(1)}$ and
$\Delta\tau^{(m)}=\tau^{(m-1)}-\tau^{(m)}$.
For a fixed $j$, Eq.~(\ref{eq:discrt}) represents a set of linear equations of the form
\beqa
\sum_{Q_2}A_{Q_1,Q_2}\cdot x_{Q_2}^{(j)}&=&b_{Q_1}^{(j)},
\label{eq:linsys}
\eeqa
where
\beqa
A_{Q_1,Q_2}&=&A_{(ip)(kq)}=\delta_{ik} \delta_{pq}-\frac{\Delta \tau^{(q)}}{2}Z_{ik}(\tau^{(p)},\tau^{(q)})\nonumber 
\eeqa
and the vectors $x_{Q_2}^{(j)}$, $b_{Q_1}^{(j)}$ are defined to be
\beqa
x_{Q_2}^{(j)}&=&x_{kq}^{(j)}=G_{kj}(\tau^{(q)})\nonumber\\
b_{Q_1}^{(j)}&=&b_{ip}^{(j)}=\delta_{ij}G_{i,0}(\tau^{(p)}).\nonumber
\eeqa
The self-energy $\Sigma^c$ of Eq.(\ref{eq:self-energy-c}) has the form
\beq
\Sigma_{ij}^{c}(\tau)=\Sigma_{c,ij}[G](\tau)+\delta(\tau)\left[\Sigma_{ij}^{HF}[G(0^-)]-\Sigma_{ij}^0\right],
\label{eq:simple}
\eeq
where $\Sigma^{HF}$ is the Hartree-Fock part of the self-energy defined in Eq.(\ref{eq:sigmaHF})
and $\Sigma_c [G]$ the remaining correlation part.
The convolution integral~(\ref{eq:convolution}) can therefore be simplified to
\beqa
Z_{ik}(\tau^{(p)},\tau^{(q)}) &=& G_{i,0}(\tau^{(p)}-\tau^{(q)})\left[\Sigma_{ik}^{HF}[G (0^-)]-\Sigma_{ik}^0 \right]\nonumber \\
&+& \int_{-\beta}^0 d\tau G_{i,0}(\tau^{(p)}-\tau)\Sigma_{c,ik}(\tau-\tau^{(q)}).
\label{eq:zik}
\eeqa
When we specify the explicit form of $\Sigma_c$, the solution of the Dyson equation is
reduced to a calculation of Eq.(\ref{eq:zik}) together with the linear system of equations (\ref{eq:linsys}).
What remains to be discussed is the calculation of the self-energy itself. 
This is discussed in the next section. 

\subsection{Numerical calculation of the screened potential: The product basis technique}
 \label{numericalscreened}

To calculate the self-energy we need to solve the equation for the screened interaction. 
The screened interaction has a singular time-local part representing the
bare interaction $v$. It is therefore convenient
to subtract $v$ from $W$ and to treat its contribution to the
self-energy explicitly (this is simply the exchange part of the self-energy). 
From the remaining time nonlocal part of $W$, given by
$\widetilde{W}(\br_1,\br_2; \tau)=W(\br_1,\br_2; \tau)-\delta(\tau)v(\br_1,\br_2)$,
we can calculate the correlation part of the self-energy
\beq
\Sigma_c (\br_1, \br_2; \tau)=-G (\br_1, \br_2; \tau) \widetilde{W}(\br_1, \br_2; \tau). 
\label{eq:gwim}
\eeq
After this quantity has been calculated it can then simply be added to the
Hartree-Fock part of the self-energy to obtain the full self-energy $\Sigma [G]$.
The time-nonlocal part $\widetilde{W}$ of the screened interaction satisfies the
equation 
\begin{eqnarray}
&&\widetilde{W}(\br_1, \br_2; \tau)=\int d\br_3\,d\br_4 v(\br_1, \br_3)P(\br_3,\br_4; \tau) v(\br_4, \br_2) + \nonumber \\
&&+\int_0^\beta d\tau' \int d\br_3 d\br_4 v(\br_1,\br_3)P(\br_3, \br_4; \tau-\tau') \widetilde{W}(\br_4, \br_2; \tau') 
\label{eq:wim}
\end{eqnarray}
where 
\beq
P(\br_1, \br_2; \tau)=2G(\br_1, \br_2; \tau)G(\br_2, \br_1; -\tau).
\label{eq:pim}
\eeq
The factor of $2$ in this expression results from spin-integrations in the equation of $W$ using the
form of the Green function of Eq.(\ref{eq:Green}).
We now insert into Eq.(\ref{eq:pim}) the basis set expansion for the Green function of Eq.(\ref{eq:Greenbasis}), to obtain 
\beq
P(\br_1, \br_2; \tau)=\sum_{ijkl}P_{ijkl}(\tau)\phi_i(\br_1)\phi^*_j(\br_2)\phi_k(\br_2)\phi^*_l(\br_1)
\label{eq:ap}
\eeq
where $P_{ijkl}=2G_{ij}(\tau)G_{kl}(-\tau)$. By defining the two-electron integrals
\beqa
\widetilde{W}_{pqrs}(\tau) &=& \int d\br_1\,d\br_2\phi_p^*(\br_1)\phi_q^*(\br_2)\widetilde{W}(\br_1, \br_2;\tau)\phi_r(\br_2)\phi_s(\br_1) \nonumber \\
v_{pqrs} &=& \int d\br_1\,d\br_2\phi_p^*(\br_1)\phi_q^*(\br_2) v(\br_1, \br_2 )\phi_r(\br_2)\phi_s(\br_1) \nonumber
\eeqa
we transform Eq.(\ref{eq:wim}) into the equation
\beqa
&\widetilde{W}_{pqrs}&(\tau)=\sum_{ijkl} v_{plis} P_{ijkl}(\tau) v_{jqrk}+\nonumber \\
&&\sum_{ijkl} \int_0^\beta d\tau' v_{plis} P_{ijkl}(\tau-\tau')\widetilde{W}_{jqrk}(\tau'). 
\eeqa
If we use the multi-indices $Q_1 =(ps)$, $Q_2=(rq)$, $Q_3 = (il)$ and $Q_4=(jk)$,
then we can write this equation in a more convenient form as
\beqa
\lefteqn{ \widetilde{W}_{Q_1 Q_2} (\tau) =  \sum_{Q_3 Q_4} 
v_{Q_1 Q_3} P_{Q_3 Q_4} (\tau) v_{Q_4 Q_2} + } \nonumber \\
&& \sum_{Q_3 Q_4}  \int_{0}^{\beta} d\tau' \,  v_{Q_1 Q_3} 
P_{Q_3 Q_4}(\tau-\tau') \widetilde{W}_{Q_4 Q_2} (\tau'),
\eeqa
where we defined $v_{il,kj}= v_{ijkl}$ and similarly for $\widetilde{W}$ and $P_{il,jk}=P_{ijkl}$.
We have now obtained an equation which we can solve with the same algorithm we used 
for the Dyson equation. 

Note that in this case we effectively use a product basis $f_q (\br)=\phi_i (\br) \phi_j^* (\br)$, where 
$q=(ij)$ is a multi-index. This product basis is nonorthogonal and its size is in general much larger than we need 
in practice due to linear dependencies. We thus follow a technique developed by Aryasetiawan and 
Gunnarsson \cite{aryasetiawan94}, which allows to reduce significantly the size of the product basis $\{ f_q(\br) \}$
and the computational cost.\\
The overlap matrix $S$ for the set of orbitals $f_q (\br)$
\beq
S_{qq'}= \braket{f_q}{f_{q'}},
\label{eq:overlapmatrix}
\eeq
is diagonalized by a unitary matrix $U$
\beq
\sum_{q_1 q_2}U_{q q_1}^{\dagger} \braket{f_{q_{1}}}{f_{q_{2}}}U_{q_2 q'}=\sigma_q\delta_{qq'},
\eeq
where the eigenvalues $\sigma_q$ are positive since $S$ is a positive definite matrix.
We now define a new set of orthonormal orbitals $g_q$ as
\beq
g_q (\br)=\frac{1}{\sqrt{\sigma_q}}\sum_{q'}U_{q'q} f_{q'} (\br),
\label{eq:newbasis}
\eeq
with $\braket{g_q}{g_{q'}}=\delta_{qq'}$.
Our strategy is use the orbitals $g_q$ as a new basis and discard the functions that correspond to 
 $\sigma_q < \epsilon$ (we used $\epsilon=10^{-6}$). This leads to a much reduced basis as
compared to the set of all functions $f_q$. As described in Ref.~\onlinecite{aryasetiawan94},
this corresponds to discarding functions that are nearly linearly dependent and contribute little in the expansion.
The quantities $\Sigma$, $\widetilde{W}$ and $P$ will be represented in this new basis using
\beq
f_q (\br) = \sum_{q'} g_{q'} (\br ) \, \sqrt{\sigma_{q'}}U^\dagger_{q' q}.
\label{eq:f}
\eeq
For the irreducible polarization we then find from Eq.~(\ref{eq:ap}) that 
\beqa
\lefteqn{ P(\br_1,\br_2; \tau)=\sum_{qq'}P_{qq'}(\tau)f_{q}(\br_1)f^*_{q'}(\br_2)=}  \nonumber \\
&=&\sum_{q_1 q_2}\left [\;\sum_{q q'} U_{q_1q}^\dagger P_{qq'}(\tau)U_{q'q_2} \;\right ]\sqrt{\sigma_{q_1}\sigma_{q_2}}g_{q_1}(\br_1)g^*_{q_2}(\br_2), 
\eeqa
where 
\beq
P_{qq'} (\tau) = 2 G_{ij}(\tau) G_{kl}(-\tau).
\eeq
With $q=(il)$ and $q'=(jk)$ we have
\beq
P(\br_1,\br_2; \tau)=\sum_{q_1q_2} \widetilde{P}_{q_1q_2} g_{q_1}(\br_1)g^*_{q_2}(\br_2),
\label{eq:pshorthand}
\eeq
where 
\beq
{\widetilde{P}}_{q_1q_2}=\left [\sqrt{\sigma} U^\dagger P(\tau)U \sqrt{\sigma}\right ]_{q_1 q_2}, 
\label{eq:Ptransform}
\eeq
and $\sqrt{\sigma}$ is the diagonal matrix  $(\sqrt{\sigma})_{pq}=\delta_{pq}\sqrt{\sigma_q}$.
To calculate the screened potential
we now insert Eq.(\ref{eq:pshorthand}) into Eq.(\ref{eq:wim})
and readily obtain the matrix product
\beq
\widetilde{W}_{qq'}(\tau)=\left [v \widetilde{P}(\tau)v \right ]_{qq'} + \left [v \widetilde{P}(\tau-\tau')\widetilde{W}(\tau') \right ]_{qq'},
\label{eq:Wnewbasis}
\eeq
where we defined the matrices
\beq
\widetilde{W}_{qq'} = \int d^3 \br_1 d^3\br_2 \, g_q^* (\br_1) \widetilde{W} (\br_1, \br_2; \tau) g_{q'} (\br_2)
\label{eq:Wnewbasis2}
\eeq
and 
\beq
v_{qq'} = \int d^3\br_1  d^3 \br_2 \, g_q^* (\br_1) v(\br_1 , \br_2) g_{q'} (\br_2).
\label{eq:vnewbasis}
\eeq
It is important to note that in Eq.(\ref{eq:pshorthand}) and Eq.(\ref{eq:Wnewbasis}) the summation only runs
over the indices $q$ for which $\sigma_q > \epsilon$. We see from Eq.(\ref{eq:Ptransform})
that terms with $\sigma_q < \epsilon$ contribute little to the total sum. 
This leads to a considerable reduction of the number 
of matrix elements for $v$, $P$ and $\widetilde{W}$. Finally the correlation part of the self-energy of Eq.(\ref{eq:gwim}) is given by
\beqa
\Sigma_{c,ij}(\tau)&=&\int d^3 \br_1 \int d^3\br_2 \phi_i^*(\br_1) \Sigma_c (\br_1,\br_2; \tau)\phi_j(\br_2)  \nonumber \\
&=& -\sum_{kl}G_{kl}(\tau)\sum_{pq}\widetilde{W}_{pq}(\tau)\int d^3 \br_1 \phi_i^* (\br_1)\phi_k (\br_1)g_p(\br_1)
 \nonumber \\
&&  \times \int d^3 \br_2  \phi_j (\br_2)\phi_l^* (\br_2)g_q^*(\br_2)  \nonumber \\
&=&-\sum_{kl}G_{kl}(\tau)Z_{ik,jl},
\label{eq:newselfenergy}
\eeqa
where 
\beq
Z_{ik,jl}=\sum_{pq}\sqrt{\sigma_p}U_{ik,p}\widetilde{W}_{pq}(\tau)U^{\dagger}_{q, jl}\sqrt{\sigma_q}.
\label{eq:zikjl}
\eeq
We can summarize our procedure as follows:
in the first step the overlap matrix $S_{qq'}$ of Eq.(\ref{eq:overlapmatrix}) is obtained
and diagonalized. Further, using and Eq.(\ref{eq:newbasis}) and (\ref{eq:vnewbasis}) the
two-electron integrals in the new basis $v_{pq}$ are constructed for $p$ and $q$ such that
$\sigma_p, \sigma_q > \epsilon$. Subsequently, for the same values of $p$ and $q$ the matrix
$\widetilde{P}_{pq}(\tau)$ is constructed from Eq.(\ref{eq:Ptransform}) and
$\widetilde{W}_{pq}(\tau)$ is solved from Eq.(\ref{eq:Wnewbasis}). 
In the last step, the matrix~(\ref{eq:zikjl}) is obtained and the self-energy is calculated from 
Eq.(\ref{eq:newselfenergy}) and further used in the solution of the Dyson equation.

\section{Results}
\label{results}

\begin{figure}
\includegraphics[width=8.6cm]{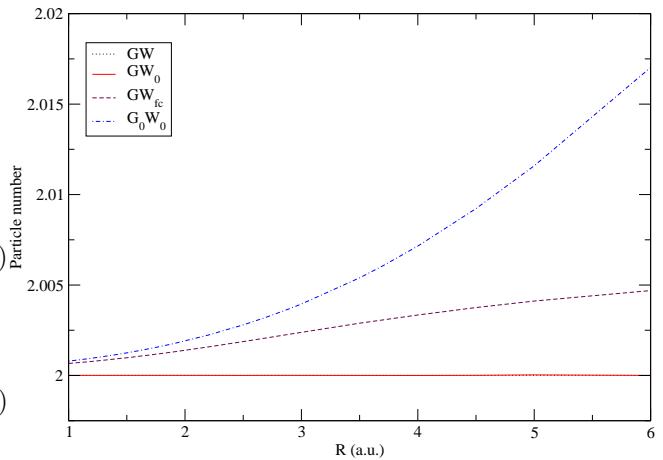} 
\caption{Particle number for H$_2$ at different interatomic distances within the SC-$GW$,  
$GW_0$, $GW_{\textrm{fc}}$ and $G_0W_0$
approximations. }
\label{fig:nop} 
\end{figure}

The various $GW$ schemes described in section~\ref{SGW-LA} are applied to a set of atoms and
diatomic molecules using the computational method of section~\ref{computational}.
Details on the basis sets are provided in Ref.~\cite{epaps}. 
In general we found that, in single processor calculations, 
the computational cost of the $GW_{\textrm{fc}}$ method is comparable
to that of the $G_0 W_0$ method, and roughly twice as fast as the $G W_0$ method.  The 
the fully self-consistent $GW$ calculations were the most time-consuming.\\
{\em Particle number conservation}. 
We start by investigating the number conservation property of the different $GW$ schemes. 
In Fig.~\ref{fig:nop} we display the particle number obtained from the trace of the Green function
for the case of the hydrogen molecule $\textrm{H}_2$ for different separations of
the nuclei. We display results for the case of SC-$GW$, 
$GW_0$, $GW_{\textrm{fc}}$ and $G_0W_0$, in which the reference Green function $G_0$ is obtained 
from a Hartree-Fock calculation. 
We see that the SC-$GW$ and $GW_0$ schemes yield an integer particle number of $N=2$
for all internuclear separations.
This is a consequence of the number conserving property of both approximations. 
This can be seen as follows. If we would
adiabatically
switch-on the two-particle interactions from zero to full coupling strength within a conserving scheme
then the particle number would be conserved during the switching. 
This is because the conserving property is independent of the strenght of the interaction and follows from
the structure of the $\Phi$-functional only.
Therefore the particle number of the final correlated state will be the same as
the particle number of the initially noninteracting system.
Hence conserving schemes always yield integer particle number for finite
systems at zero temperature.
For the case of the hydrogen molecule this is $N=2$ for all
bond distances.
For the case of $G_0 W_0$ we see that the particle number conservation is violated
as the particle number deviates from $N=2$ for all bond distances, the largest deviations
occuring for the larger bond distances. For the larger separations left-right correlation~\cite{baerendsgritsenko97}
in the hydrogen molecule, not incorporated in the Hartree-Fock part of the self-energy,
become increasingly important.
This puts more demands on the quality of the correlation part of the
self-energy and consequently nonconservation of the particle number becomes more apparent
at longer bond distances.
Although the violation
seems small (about $0.01$ electron at $R=4.5$) it should
be emphasized that a change in particle number of $0.05$ can give large changes in the spectral
features and conductive properties for molecules attached to leads. A clear example of this is
presented in the work of Thygesen~\cite{thygesen08PRL}. 
For the $GW_{\textrm{fc}}$ (See sec.~\ref{ssec:gwfc}) we also observe a violation
of the number conservation law with increasing error for larger internuclear separations.
The error with respect to $G_0W_0$ is however reduced by a factor of $3$ at $R=5.5$
as a consequence of a partial inclusion of self-consistency.

{\em Ground state energies}. For the various $GW$ schemes of section~\ref{SGW-LA} we calculated
the total energies of some atoms and diatomic molecules from Eq.(\ref{eq:uxc}).
The reference Green function $G_0$ for the nonself-consistent schemes was obtained 
from a Hartree-Fock calculation. 
In Table~\ref{tab:totale} we show the results. 
From comparison with benchmark configuration interaction (CI) results we see that
the total energies of atoms and molecules calculated within all schemes are not very accurate.
However, as we will see later, energy differences are much better produced.
We can nevertheless make a number of useful observations from the total energies.
We first note that all approximations produce a total energy that is lower than the 
benchmark CI result, with the $G_0 W_0$ generally producing the lowest
and thereby the worst values.
Both the $GW_0$ and the $GW_{\textrm{fc}}$ methods yield total energies in excellent agreement
with SC-$GW$ results, where for most systems the difference is $10^{-3}$ Hartree or less. 
This means that both the $GW_0$ and the $GW_{\textrm{fc}}$ methods can be used to
make an accurate prediction for the SC-$GW$ energy at a much lower computational cost
than the fully self-consistent calculation.
\begin{table*}
\caption{Total energies (in Hartrees) calculated from the $GW$ approximation at various levels of self-consistency
compared to CI values.}
\label{tab:totale}
\begin{ruledtabular}
\begin{tabular}{lcccccr}
System & $E^{G_0W_0}[G_{\rm {HF}}]$ & $E^{GW_0}[G_{\rm {HF}}]$ & $E^{GW_{\textrm{fc}}} [G_{\rm {HF}}]$ &  $E_{\rm {SC}}^{GW}$ &CI \\  
\hline
He        &   -2.9354&   -2.9271 & -2.9277 & -2.9278 & -2.9037$^1$ \\
Be        &  -14.7405&  -14.6882 & -14.7032  & -14.7024 &  -14.6674$^1$ \\
Be$^{2+}$ & -13.6929& -13.6886 & -13.6887  & -13.6885&  -13.6556$^1$\\
Ne        & -129.0885& -129.0517 & -129.0506  & -129.0499&  -128.9376$^1$\\
Mg        & -200.2924& -200.1759 & -200.1775  & -200.1762&  -200.053$^1$\\
Mg$^{2+}$ & -199.3785& -199.3451 & -199.3454  & -199.3457&  -199.2204$^1$\\
H$_2$     &  -1.1985&  -1.1889 & -1.1891   & -1.1887& -1.133$^2$\\
LiH       &   -8.1113&   -8.0999 & -8.0997  & -8.0995& -8.040$^3$\\
\end{tabular}
\end{ruledtabular}
\begin{flushleft}
$^1$From Ref. \cite{chakravorty93}. $^2$From Ref. \cite{vanleeuwen94}.
$^3$From Ref. \cite{li03}.\\
\end{flushleft}
\end{table*}

\begin{figure}
\includegraphics[width=8.6cm]{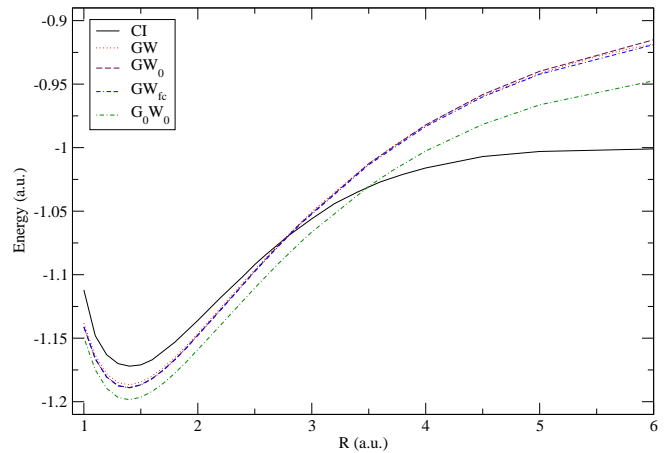} 
\caption{ The total energy of the H$_2$ molecule, as a function of the interatomic distance, calculated
from the $GW$ approximation at various levels of self-consistency and CI \cite{vanleeuwen94}.}
\label{fig:gwss}
\end{figure}

{\em Binding curve}.
The calculation of binding curves is a good test for the quality
of total energy calculations.
In Fig.~\ref{fig:gwss} we display the binding curve of the  H$_2$ molecule
for the various $GW$ schemes together with benchmark CI results. 
The reference Green function $G_0$ was taken from a Hartree-Fock calculation. 
We further checked that using a $G_0$ obtained from an LDA calculation only
influences the results slightly. 
For the values of the energies around the bond minimum we see the same trend 
that we observed before:  all $GW$ schemes
lead to a total energy that is lower than the benchmark CI results with $G_0W_0$
being the lowest. The total energies of the partially self-consistent schemes $GW_0$
and $GW_{\textrm{fc}}$ are very close to the fully self-consistent $GW$
results for all bond distances.
Although all $GW$ schemes considerably improve the bonding curve obtained
from an uncorrelated Hartree-Fock calculation it is clear that all these schemes deviate
considerably from the CI results in the infinite atomic separation limit.
To cure this feature one either has to do a spin-polarized calculation
or go beyond the $GW$ approximation and include vertex diagrams in the 
diagrammatic expansion for the self-energy.
The shape of the binding curve around the bond minimum is well
reproduced by the SC-$GW$, $GW_0$ and $GW_{fc}$ schemes, implying that these
methods may be used to obtain accurate vibrational frequencies.
Since the shape of the bonding curve is only determined by total energy 
differences, this already indicates that these approximations may perform 
better in obtaining the energy differences than in obtaining total
energies. 

{\em Two-electron removal energies}.
To test the performance of the various $GW$ schemes in obtaining energy differences,
we investigated the two-electron removal energies of the beryllium and magnesium atom.
Since these atoms and their doubly ionized counterparts are closed shell they were
suitable test systems. Moreover, the beryllium atom is a well-known case for which
electron correlations play an important role due to strong mixing of the $2s$  and
$2p$ states in a configuration expansion.
In table~\ref{tab:subtract}, we display the two-electron removal energies for
various $GW$ schemes as well as for the Hartree-Fock approximation.
The reference Green function $G_0$ is again obtained from a Hartree-Fock calculation.
The self-consistent and partially self-consistent $GW$ schemes yield results within 
$0.1$ eV from the experimental values and
considerable improve the HF values that differ with more than 1 eV from experiment.
The $G_0W_0$ approximation does not improve at all on the HF approximation and gives considerably
worse results than the other $GW$ schemes.
We further see that both the $GW_0$ and the $GW_{\textrm{fc}}$ approximations give removal energies
that are in excellent agreement with the fully self-consistent $GW$ results.

\begin{table}
\caption {Two-electron removal energies $E_{N-2}-E_{N}$ (in eV) calculated from the Hartree-Fock and from
the $GW$ approximation at various levels of self-consistency, compared to the experimental values.}
\label{tab:subtract}
\begin{ruledtabular}
\begin{tabular}{lcccccc}
System & HF & $\Delta E^{G_0W_0}$ & $\Delta E^{GW_0}$ & $\Delta E^{GW_{\textrm{fc}}}$ & $\Delta E^{GW}_{SC}$  & Expt.$^1$ \\ 
\hline
Mg -  Mg$^{2+}$ & 21.33 & 24.86 & 22.61 & 22.64 & 22.59 & 22.68 \\
Be - Be$^{2+}$ & 26.17 & 28.50 & 27.20 & 27.61 & 27.59 & 27.53 \\
\end{tabular}
\end{ruledtabular}
\begin{flushleft}
$^1$From Ref. \cite {LLK}.
\end{flushleft}
\end{table}
 
{\em Ionization Potentials}. In Table~\ref{tab:ioniz} we show the ionization potentials 
obtained with the various $GW$ methods for a number of atoms and diatomic molecules.
These ionization potentials were obtained using the extended Koopmans theorem, as explained in Appendix~\ref{A0_EKT}.
For $G_0W_0$ the results shown in the first column were obtained by using a reference Green
function $G_0$ from a local density functional (LDA) calculation
using the parametrization of the exchange-correlation functional due to Vosko et al.~\cite{voskoLDA}. 
In all other cases we used a reference Green
function from a Hartree-Fock calculation. 
We see that the ionization potentials of fully self-consistent $GW$ agree well with
the experimental values, the main exceptions being the H$_2$ molecule and the Be atom,
which show a deviation of respectively $0.8$ and $0.5$ eV.
The other partially self-consistent approaches $G W_0$ and $GW_{\textrm{fc}}$ yield
results that are very close to the fully self-consistent results.
The $G_0W_0$ approximation based on the LDA reference Green function performs a bit worse than
the self-consistent $GW$ scheme. For He and LiH there is an error of about 1 eV
and for Ne and H$_2$ an error of about $0.5$ eV. Performing a $G_0 W_0$
calculation based on a HF reference $G_0$ instead improves the results for several systems
but worsens the agreement for H$_2$ which is $1.1$ eV in error.
The dependence on the reference Green function $G_0$ within the
$G_0W_0$ method is clearly unsatisfactory.
The partially self-consistent approximations suffer much less from this problem.
For those schemes we found that changing the
reference Green function from a HF one to an LDA one, only slightly changes
the results.
\begin{table}
\caption{ Ionization potentials (eV) calculated from the extended Koopmans theorem
from various $GW$ approaches.}
\label{tab:ioniz}
\begin{ruledtabular}
\begin{tabular}{lcccccc}
Sys. & $G_0^{(LDA)}W_0 $ & $G_0^{(HF)}W_0 $ & $GW_0$ &  $GW_{fc}$ & SC-GW & Expt.$^1$\\  
\hline
 He        &   23.65    & 24.75 &  24.59 & 24.56 & 24.56 & 24.59 \\
 Be        &   8.88 & 9.19 &  8.82 & 8.81 & 8.66 & 9.32 \\
 Ne        &  21.06 & 21.91 &  21.90 & 21.82 & 21.77 & 21.56 \\
 Mg        &  7.52 & 7.69 &  7.43 & 7.38 &  7.28  & 7.65\\
 H$_2$  	 & 15.92 & 16.52 &  16.31 & 16.22 &  16.22 & 15.43\\
 LiH       & 6.87 & 8.19 &  7.71 & 7.85 & 7.85 & 7.9 \\
\end{tabular}
\end{ruledtabular}
\begin{flushleft}
$^1$From Ref. \cite{LLK} \\
\end{flushleft}
\end{table}

\section{Summary and conclusions}
\label{cz}

We investigated the performance of the $GW$ at different levels of self-consistency 
for the case of atoms and diatomic molecules.
Our main motivation for studying fully self-consistent $\Phi$-derivable schemes 
was that they provide unambiguous results for different observables and the
fact that they satisfy important conservation laws that are important in future
nonequilibrium applications of the theory~\cite{myohanen08}.
We adressed the question to what extent partially self-consistent schemes 
can reproduce the results of a fully self-consistent $GW$ calculation.
We found that both the $GW_0$ method, as well as the $GW_{\textrm{fc}}$ scheme proposed by us,
yield results in close agreement with fully self-consistent $GW$ calculations.
We further checked the number conservation properties of the various
schemes. The fully self-consistent $GW$ scheme being $\Phi$-derivable does
satisfy all conservation laws, but also the partially self-consistent $GW_0$ 
approximation was shown to be number conserving. The nonself-consistent $G_0 W_0$ 
and the partially self-consistent $GW_{\textrm{fc}}$ approximations both violate the number 
conservation laws but, due to the partial self-consistency in $GW_{\textrm{fc}}$, the errors are
much reduced in this scheme. A major advantage of the latter scheme is, however,
that it produces results that are close to the fully self-consistent $GW$
results at a much lower computational cost. It will therefore be very valuable
to test this method on solid state systems for which self-consistent $GW$
calculations are difficult to perform due to the large computational effort. 
In this way it will be possible to get further insight into the performance
of self-consistent $GW$ for a large class of extended systems.
Work on application of the fully self-consistent $GW$ method to transport
phenomena is in progress~\cite{myohanen08}.

\appendix 
\section{Ionization potentials from the Extended Koopmans Theorem}
\label{A0_EKT}

Here we give a brief description on the way we extract the ionization energies 
from the Green function using the extended Koopmans theorem~\cite{katrieldavidson, dayEKT75, smithEKT75, sundholmEKT93, morrison_ayersEKTBe95}. As input, this method only needs the Green function and 
its time derivative at $\tau=0^-$ on the imaginary time axis.
We define an $N-1$ particle state
\begin{eqnarray}
|\Phi^{N-1}[u_i]>&=&\int\;d\bx\, u_i(\bx)\hat{\psi}(\bx)|\Psi_0^N>,
\end{eqnarray}
where $u_i(\bx)$ is determined by requiring
the functional 
\beq
E^{N-1}[u_i]=\frac{\langle \Phi^{N-1} [u_i]|\hat{H}|\Phi^{N-1} [u_i]\rangle}
{\langle \Phi_{N-1} [u_i]|\Phi_{N-1} [u_i]\rangle},
\eeq 
which describes the energy of the $N-1$ particle system, to be stationary with respect to variations in $u_i$.
This amounts to minimizing the energy of the $N-1$ system
by choosing an optimal value for $u_i$. 
We find
\beqa
\int d\bx  \bra{\psin}\hpsid(\bxp) \big[\psih(\bx),\hat{H}\big] \ket{\psin} u_i(\bx) = \nonumber\\
(E_{0}^{N}-E^{N-1}_i)\int d\bx \bra{\psin}\hpsid(\bxp) \psih(\bx)\ket{\psin} u_i(\bx),
\label{eq:1ekt}
\eeqa
where the last term contains the density matrix.
This quantity is easily obtained from the Green function as
   \beq
   \rho(\bx,\bx')= \langle \psi_0^N| \hat{\psi}_H^\dagger(\bx'\tau)\hat{\psi}_H(\bx \tau) |\Psi_0^N \rangle =
 \lim_{\eta\rightarrow 0} G(\bx,\bx',-\eta)
   \label{eq:1partdensmatrix}
\eeq
 {\it i.e.} $\rho(\bx,\bxp)=\tilde{G}(\bx,\bxp;0^-)$ or $\rho_{ij}=G_{ij}(0^-)$ in
molecular orbital basis~\cite{dahlen05b}. Also the expectation value under the integral on the righthand side of Eq.(\ref{eq:1ekt}),
is easily obtained from the Green function
\beqa
-\partial_\tau G(\bx,\bxp;\tau)\arrowvert_{\tau={0^-}}&=&\bra{\psin}\psihd(\bxp)\big[\psih(\bx),\hat{H}\big]\ket{\psin} \nonumber \\
&=& \Delta(\bx,\bxp).
\label{eq:2ekt}
\eeqa
In this derivation we used a zero-temperature formulation but making a connection to the finite temperature formalism
is straightforward. 
When we take into account that, in the finite temperature formalism, we
included the chemical potential in the one-body part of the Hamiltonian (see Eq.(\ref{eq:h0}),
then from (\ref{eq:1ekt}) and (\ref{eq:2ekt}) we obtain the eigenvalue equation
\begin{eqnarray}
&&\int d\bx \; \Delta(\bx,\bxp)u_i(\bx)=\nonumber \\ 
&&= (E_0^N-E_i^{N-1}-\mu) \int d\bx \;\rho(\bx,\bxp)u_i(\bx),
\label{eq:EKT}
\end{eqnarray}
where $\rho$ and $\Delta$ are calculated according to Eq.(\ref{eq:1partdensmatrix}, \ref{eq:2ekt}).
A similar equation for the electron affinities can similarly be derived starting from an $N+1$-state.
Since both matrices $\rho$ and $\Delta$ are easily evaluated from the Green function,
Eq.(\ref{eq:EKT}) provides an easy way to extract removal energies from knowledge of
the Green function on the imaginary time axis.

For completeness we mention that the extended Koopmans method also provides a simple way to extract
quasiparticle or Dyson orbitals~\cite{morrison_ayersEKTBe95} and to construct the Green function on the
real frequency axis.
The Dyson orbitals are given by
\beqa
f_i(\bx)&=&\bra{\Phi_i^{N-1}}\hat{\psi}(\bx)\ket{\Psi_0^N}= \nonumber \\
&=&\int d\bx' \; u_i^*(\bx') \bra{\Psi^{N}} \hat{\psi}^\dagger(\bx') \hat{\psi}(\bx)\ket{\Psi_0^N}=\nonumber \\
&=&\int d\bx' \;\rho(\bx,\bx')u_i^* (\bx').
\eeqa
In terms of these orbitals and the extended Koopmans eigenvalues the hole-part of the Green function
is then given on the real frequency axis as
\beq
G(\bx,\bxp;\omega)= \sum_n \frac{f_n(\bx)f_n^*(\bxp)}{\omega-(E_0^N-E_n^{N-1}+\mu)+i\eta}.\nonumber
\eeq
Similar derivations can be carried out for the affinities and the corresponding Dyson orbitals
from which the particle-part of the Green function can be constructed on the real axis. 

\section{The Uniform Power Mesh}
\label{A2}

The uniform power mesh (UPM)~\cite{ku02} is a one-dimensional grid on an interval $[0,\beta ]$
which becomes more dense at the endpoints. Therefore, it is well-suited to describe the Green function 
on the imaginary time axis, since it behaves exponentially around $\tau=0$ and $\tau=\pm\beta$ 
\cite{dahlen05b, ku02}. The UPM is defined by two integers $u$ and $p$ and the length of the interval $\beta$. 
The procedure to construct it is simple: we consider the $2(p-1)$ intervals
$[0,\beta_j]$ and $[\beta-\beta_j,\beta]$ for $j=1,\ldots,p-1$ with $\beta_j=\beta/2^j$, and
divide each of these intervals in $2u$ subintervals of equal lenght. 
The endpoints of all these intervals define our grid which has
$2pu+1$ grid points.


\end{document}